\documentclass[conference]{IEEEtran}

\ifCLASSINFOpdf
\else
\fi
\usepackage{graphicx}
\usepackage{tabularx,booktabs}
\usepackage{dingbat}
\usepackage{diagbox}
\usepackage{multirow} 
\usepackage[hyphens]{url}
\usepackage[hidelinks,breaklinks]{hyperref}
\usepackage{xcolor}
\usepackage{amsfonts, amsmath, amsthm, amssymb}
\usepackage{subcaption}
\hypersetup{breaklinks=true}
\usepackage{tikz}
\usepackage{textcomp}
\usepackage{lipsum}
\usepackage{setspace}
\usepackage{svg}

\IEEEoverridecommandlockouts
\newcommand\copyrighttext{%
  \footnotesize \textcopyright 2025 IEEE.  Personal use of this material is permitted.  Permission from IEEE must be obtained for all other uses, in any current or future media, including reprinting/republishing this material for advertising or promotional purposes, creating new collective works, for resale or redistribution to servers or lists, or reuse of any copyrighted component of this work in other works.}
\newcommand\copyrightnotice{%
\begin{tikzpicture}[remember picture,overlay]
\node[anchor=south,yshift=10pt] at (current page.south) {\fbox{\parbox{\dimexpr\textwidth-\fboxsep-\fboxrule\relax}{\copyrighttext}}};
\end{tikzpicture}%
}

\usepackage{graphicx}
\graphicspath{{figures/}}
\begin{document}

\bstctlcite{IEEEexample:BSTcontrol}
%
\title{Performance Analysis of 5G FR2 (mmWave) Downlink 256QAM on Commercial 5G Networks}

\author{
    \IEEEauthorblockN{Kasidis Arunruangsirilert\IEEEauthorrefmark{1}\IEEEauthorrefmark{3}, Pasapong Wongprasert\IEEEauthorrefmark{2}\IEEEauthorrefmark{3}, Jiro Katto\IEEEauthorrefmark{1}}
    \IEEEauthorblockA{\IEEEauthorrefmark{1}Department of Computer Science and Communications Engineering, Waseda University, Tokyo, Japan}
    \IEEEauthorblockA{\IEEEauthorrefmark{2}Department of Electrical Engineering, Chulalongkorn University, Bangkok, Thailand
    \\\{kasidis, katto\}@katto.comm.waseda.ac.jp, 6670331021@student.chula.ac.th}
    \IEEEauthorblockA{\IEEEauthorrefmark{3}Authors contributed equally to this research.}
    
}
\vspace{-2mm}
%
\maketitle

\vspace{-2mm}
\begin{abstract}

 \setstretch{0.93}
The 5G New Radio (NR) standard introduces new frequency bands allocated in Frequency Range 2 (FR2) to support enhanced Mobile Broadband (eMBB) in congested environments and enables new use cases such as Ultra-Reliable Low Latency Communication (URLLC). The 3GPP introduced 256QAM support for FR2 frequency bands to further enhance downlink capacity. However, sustaining 256QAM on FR2 in practical environments is challenging due to strong path loss and susceptibility to distortion. While 256QAM can improve theoretical throughput by 33\%, compared to 64QAM, and is widely adopted in FR1, its real-world impact when utilized in FR2 is questionable, given the significant path loss and distortions experienced in the FR2 range. Additionally, using higher modulation correlates to higher BLER, increased instability, and retransmission. Moreover, 256QAM also utilizes a different MCS table defining the modulation and code rate at different Channel Quality Indexes (CQI), affecting the UE's link adaptation behavior. This paper investigates the real-world performance of 256QAM utilization on FR2 bands in two countries, across three RAN manufacturers, and in both NSA (EN-DC) and SA (NR-DC) configurations, under various scenarios, including open-air plazas, city centers, footbridges, train station platforms, and stationary environments. The results show that 256QAM provides a reasonable throughput gain when stationary but marginal improvements when there is UE mobility while increasing the probability of NACK responses, increasing BLER, and the number of retransmissions. Finally, MATLAB simulations are run to validate the findings as well as explore the effect of the recently introduced 1024QAM on FR2.
\end{abstract}


 \copyrightnotice

%
\IEEEpeerreviewmaketitle

\vspace{-2mm}
\section{Introduction}

One of the primary objectives of 5G New Radio (NR) is to achieve peak downlink throughput of up to 20 Gbps \cite{itu_m2083}. To facilitate this, Frequency Range 2 (FR2), also known as millimeter wave (mmWave), has been introduced, enabling the utilization of frequencies above 24 GHz—including the K-band and Ka-band—for mobile communications. The substantial spectrum available in this range allows Mobile Network Operators (MNOs) worldwide to deliver high throughput to user equipment (UE). However, this advantage comes at the expense of coverage and reliability due to significant path loss at higher carrier frequencies \cite{8891537}. Consequently, mmWave primarily serves as a capacity layer, deployed as small cells to supplement capacity in areas with high user density, thereby relieving the load on lower frequency bands and preserving resources on these coverage layers for UEs at the cell edges.

As the world progresses toward the 6G era, an increasing number of downlink-intensive use cases have emerged, including IP multicasting for UHD TV broadcasting over 5G \cite{7329950,tsuchida_2019}, Fixed Wireless Access (FWA) \cite{8484810,9628442}, and the transmission of large multimedia data for AI-related tasks. Despite this surge in demand, the frequency spectrum remains a limited resource, making the maximization of spectral efficiency essential. Similar to 4G Long Term Evolution (LTE) and 5G Frequency Range 1 (FR1) or Sub-6 GHz bands, the 3rd Generation Partnership Project (3GPP) proposed the application of Downlink 256QAM (DL-256QAM) on FR2 to enhance data throughput as part of their Release 15 \cite{3GPP_38-912}. Commercial UEs began implementing this feature in early 2022 with the release of the Japanese versions of the Samsung Galaxy S22 Ultra and Sony Xperia 1 IV, featuring the Qualcomm Snapdragon X65 RF Modem System. Although 256QAM modulation theoretically offers up to a 33\% increase in spectral efficiency, previous research indicates that sustaining this modulation scheme in real-world deployments is challenging in both uplink and downlink directions—even in Sub-6 GHz bands—due to high Signal-to-Interference-plus-Noise Ratio (SINR) requirements \cite{7470807, 10570635}. Additionally, using higher modulation correlates to an increase in transmission errors, therefore increasing the probability of retransmissions. In 5G NR, this translates to an increase in NACK responses, prompting retransmission until ACK is received. \looseness=-1

Moreover, Radio Access Network (RAN) documentations from multiple manufacturers suggest that enabling 256QAM modulation may negatively impact UEs with weak signals, leading to reduced throughput. This is attributable to the switch from the \textit{qam64} table (MCS index table 1 for PDSCH) to the \textit{qam256} table (MCS index table 2 for PDSCH) when 256QAM is enabled. Given that the 5G Physical Downlink Shared Channel (PDSCH) MCS table contains a maximum of 32 entries, the step size between each MCS index in the \textit{qam256} table is coarser than in the \textit{qam64} counterpart, resulting in less precise link adaptation due to the omission of intermediate MCS indexes. To address this, some RAN manufacturers offer proprietary solutions. For example, in Huawei RAN systems, configuring the parameter \textit{"DlLinkAdaptAlgoSwitch"} to \textit{"DL\_MCS\_TABLE\_ADAPT\_SW"} allows the RAN to automatically switch between the \textit{qam64} and the \textit{qam256} tables based on channel quality, combining the benefits of both. Additionally, RAN manufacturers typically classify DL-256QAM as an additional feature and usually charge MNOs a license fee on a per-cell basis, so enabling 256QAM in suboptimal conditions could incur extra costs without providing any benefits. Given the significantly higher path loss of FR2 compared to typical mobile communication bands, it is important to study how 256QAM modulation performs on mmWave bands. Such an investigation can assist MNOs in network optimization by minimizing potential negative impacts and avoiding unnecessary licensing costs associated with enabling DL-256QAM in non-ideal scenarios.

This paper, therefore, thoroughly investigates the network impact of enabling DL-256QAM on FR2, focusing on downlink throughput, retransmission rate, and modulation utilization. Experiments were conducted in two countries—Japan and Thailand—across three different MNOs: SoftBank Japan, au by KDDI Japan, and Advanced Info Service (AIS) Thailand, each employing different RAN manufacturers (Ericsson, Samsung, and Huawei, respectively). The frequency ranges each MNO uses also differ, with n257 (28 GHz) in Japan and n258 (26 GHz) in Thailand. Additionally, Japanese MNOs are among the first globally to commercially deploy 5G NR Dual Connectivity (NR-DC), which allows mmWave bands to be utilized in 5G Standalone (SA) deployments by enabling Sub-6 GHz and mmWave aggregation as early as late-2022 \cite{qualcomm_2022}. This scenario provides an opportunity to examine how DL-256QAM behaves in both Evolved-Universal Terrestrial Radio Access–Dual Connectivity (EN-DC, Non-Standalone) and NR-DC (SA) deployment models, thereby highlighting differences across frequency bands and deployment strategies. Data collection was performed in various locations, including open-air plazas, city centers, footbridges, train station platforms, and indoor environments, under two mobility conditions: walking and stationary. Finally, MATLAB simulations were run in three case scenarios: stationary, walking, and biking, to validate the real-world results. Simulations of 1024QAM modulation are also carried out to understand the effect of 1024QAM, which was recently introduced in FR1 as part of 3GPP Release 17 \cite{3GPP_38-306}, on FR2. \looseness=-1


\section{Experiment Environment}

\subsection{User Equipment (UE)}

The Samsung Galaxy S22 Ultra (SC-52C), equipped with a Qualcomm Snapdragon X65 5G RF Modem \cite{qualcomm_x65}, was used as the User Equipment (UE) for the experiment conducted in Japan. In Thailand, the Sony Portable Data Transmitter (PDT-FP1), with Qualcomm Snapdragon X70 5G RF Modem \cite{qualcomm_x70}, was utilized due to the difference in the frequency band. Both UEs support downlink 256QAM and can be modified to limit the modulation capability to 64QAM for the experimental requirements through firmware modification. The modification was verified by analyzing the \textit{UECapabilityInformation} packet. \textit{Network Signal Guru (NSG)} was used to collect the modem log data on the SC-52C smartphone. For the PDT-FP1, the modem log was collected through the DIAG interface using \textit{AirScreen}. The logs were then analyzed using both \textit{AirScreen} and \textit{Qualcomm Commercial Analysis Toolkit (QCAT)}. Since 5G FR2 is computationally intensive, resulting in excessive heat generation. For long walk tests in the neighborhoods, the test UE was equipped with an 18W thermoelectric cooler to avoid the possibility of network performance drops due to thermal throttling as well as force shutdown due to device's thermal mitigation mechanism.

\vspace{-1mm}
\subsection{Network Environment}

The experiments were conducted on commercial 5G networks in Japan and Thailand, specifically on SoftBank and KDDI networks in Tokyo, Japan, and the AIS network in Bangkok, Thailand. The RAN configurations differ slightly across the three networks regarding parameters and RRC features, as outlined in Table \ref{tab_RRCConfig}. The RAN vendor also differed across these networks, with AIS using Huawei, KDDI using Samsung, and SoftBank using Ericsson. These variations in configurations and vendor equipment were intended to introduce variability in network performance, allowing for a broader analysis of FR2 or mmWave network behavior across diverse settings. During the experiment, data collection was performed during walk tests and also stationary tests to understand the performance impact when mobility is introduced.

\begin{table}[!tbp]
\vspace{1mm}
\setstretch{0.8}
\caption{Summary of Network RRC Features differences}
\vspace{-1.5mm}
\centering
\label{tab_RRCConfig}
\resizebox{6cm}{!}{\begin{tabular}{@{}lccc@{}}
\toprule
Parameter                  &AIS&KDDI&SoftBank   \\\midrule
Vendor & Huawei & Samsung & Ericsson\\
CSI-RS for TRS-info & Yes & No & Yes\\
QCL/TCI States & Yes & No & Yes\\
Meas Report& rsrp, rsrq, sinr & rsrp, rsrq& rsrp, sinr\\
\bottomrule
\end{tabular}}
\vspace{-4mm}
\end{table}

\begin{table}[!tbp]
\vspace{1mm}
\setstretch{0.8}
\caption{Summary of Simulation Parameters}
\vspace{-1.5mm}
\centering
\label{tab_SimParams}
\resizebox{5cm}{!}{\begin{tabular}{@{}lc@{}}
\toprule
Parameter                  &Value\\\midrule
Carrier Frequency& 24.8 GHz\\
Channel Bandwidth& 100 MHz\\
Number of Resource Block&66\\
Subcarrier Spacing (SCS) & 120 kHz\\
TDD Slots Configuration (DL/UL)&3/1\\
TDD Symbols Configuration (DL/UL)&10/1\\
Tx/Rx Mode & 4T4R \\
EIRP & 67 dB \\
Receiver Gain & 29.5 dB\\\midrule
UE Receive Gain & 1 dB\\
UE Tx/Rx Antenna Count & 2T2R\\
UE Transmit Power & 23 dBm\\
\bottomrule
\end{tabular}}
\vspace{-6mm}
\end{table}

\subsection{Field Test Setup}

Since 5G FR2 coverage is quite limited, the experiment location is based on the availability of 5G FR2. Attempts had been made to choose the location with a wide variety of RF conditions and deployment configurations. Overall, eight locations were chosen, in which, 11 head-to-head experiments (nine with mobility, two at rest) were conducted at five locations with ease of repeatability. However, when repeatability is deemed difficult due to reasons such as wide 5G FR2 coverage or an area being too crowded, UE was set to use \textit{qam256} table, and one trial was conducted. LTE Band 3 (1.8 GHz) was used as the anchor band for 5G NSA experiments, and for 5G SA, NR Band n77 (3.7 GHz) was used to anchor the FR2 band.

\begin{figure}[t!]
\vspace{1mm}
\centering
\begin{subfigure}{.24\textwidth}
  \centering
  \includegraphics[width=0.95\linewidth]{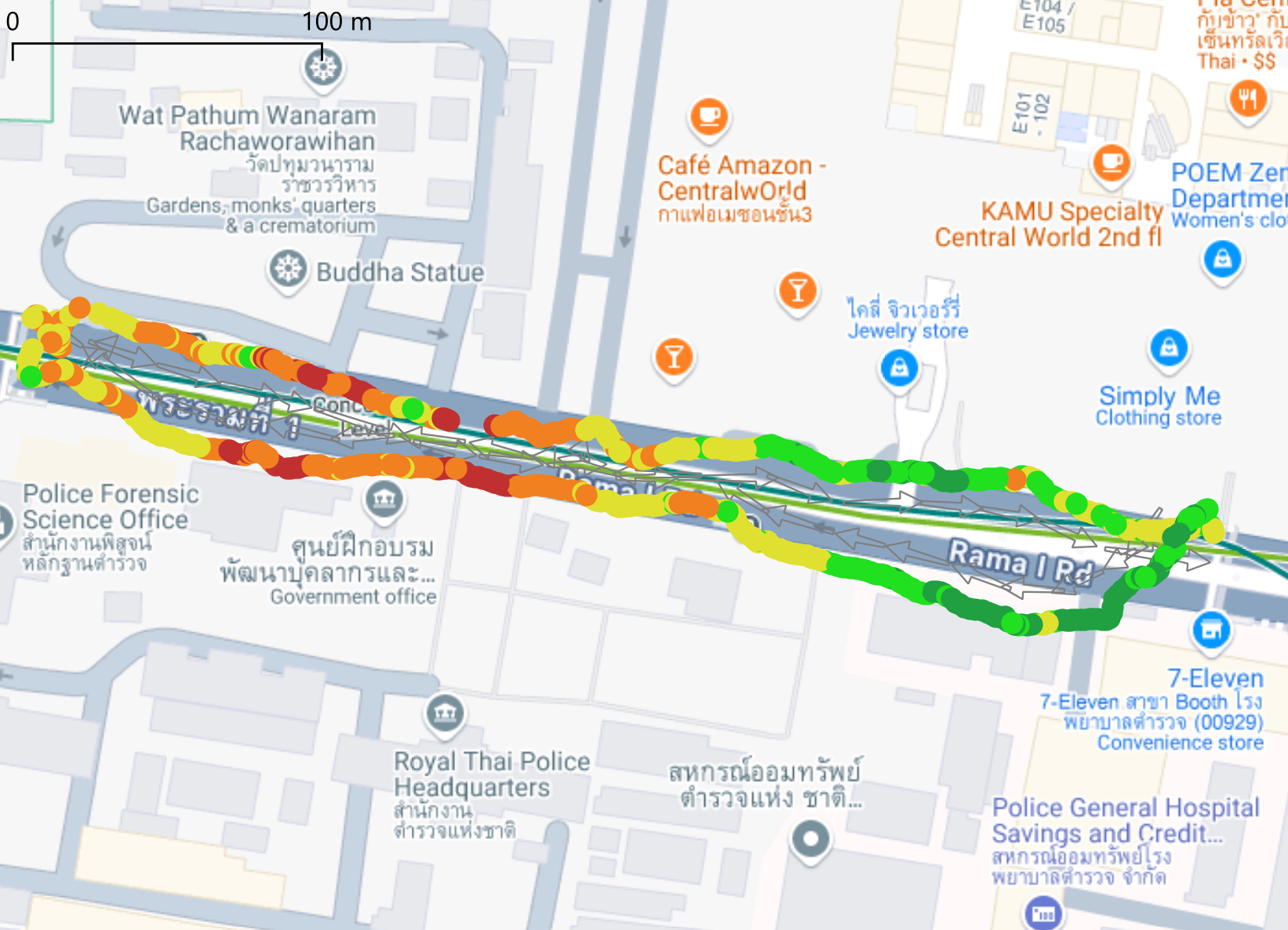}
  \vspace{-1mm}
  \caption{Sky Bridge (TH)}
  \label{fig:AIS_Walk}
\end{subfigure}%
\begin{subfigure}{.24\textwidth}
  \centering
  \includegraphics[width=0.95\linewidth]{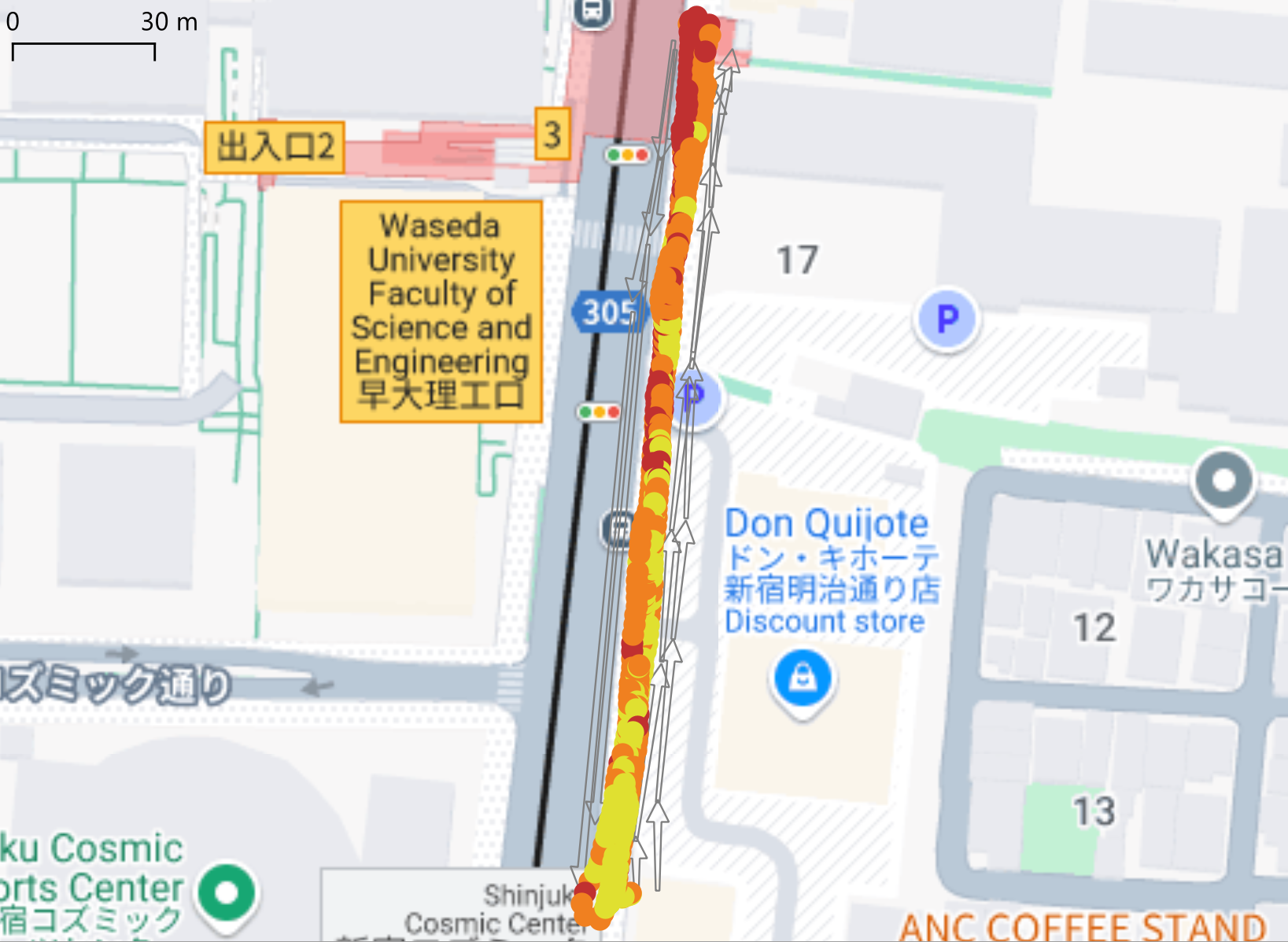}
  \vspace{-1mm}
  \caption{Waseda Sidewalk (JP)}
  \label{fig:Waseda_Walk}
\end{subfigure}\\
\begin{subfigure}{.24\textwidth}
  \centering
  \includegraphics[width=0.95\linewidth]{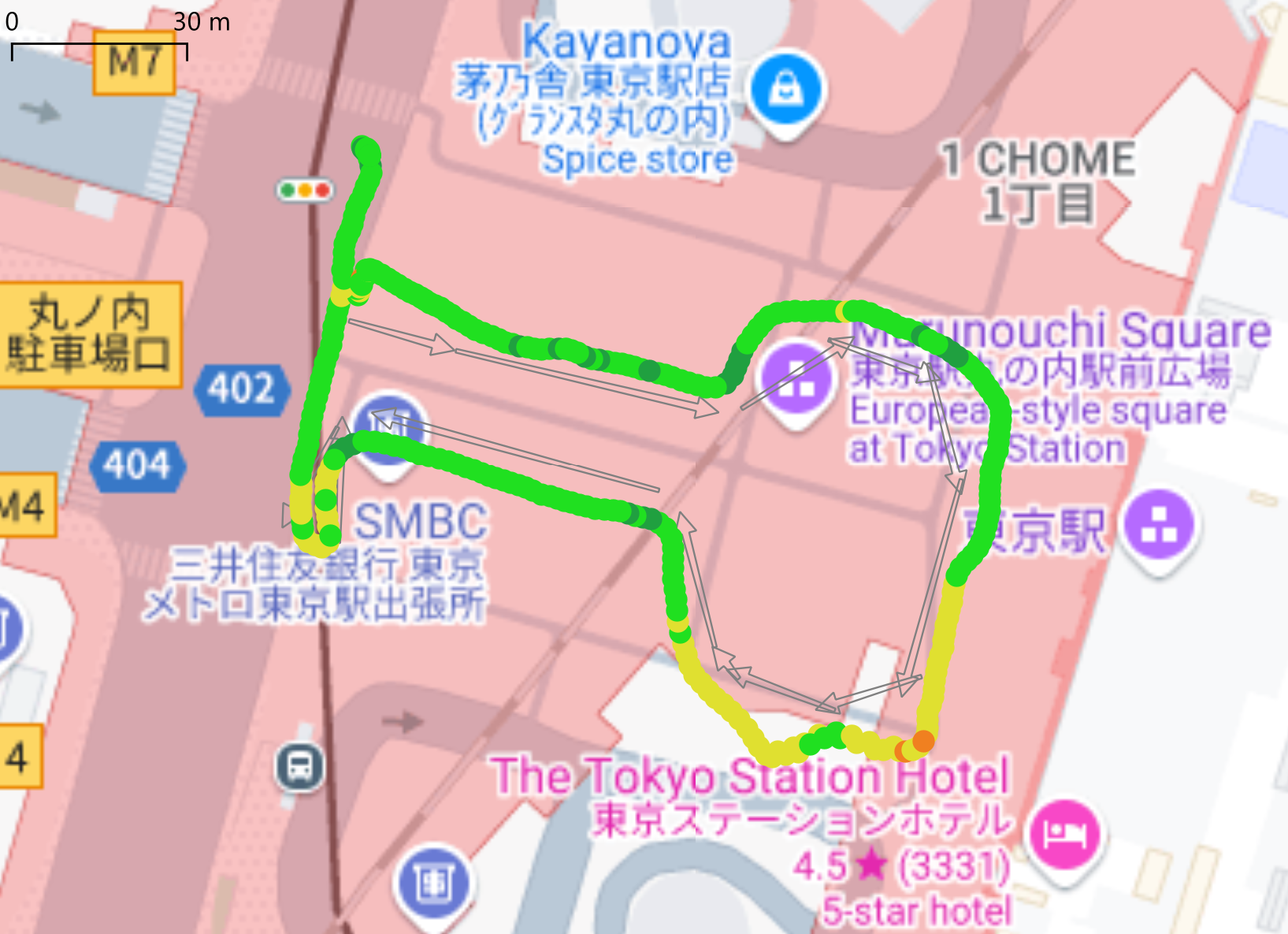}
  \vspace{-1mm}
  \caption{Tokyo Station Square (JP)}
  \label{fig:TokyoStation}
\end{subfigure}%
\begin{subfigure}{.24\textwidth}
  \centering
  \includegraphics[width=0.95\linewidth]{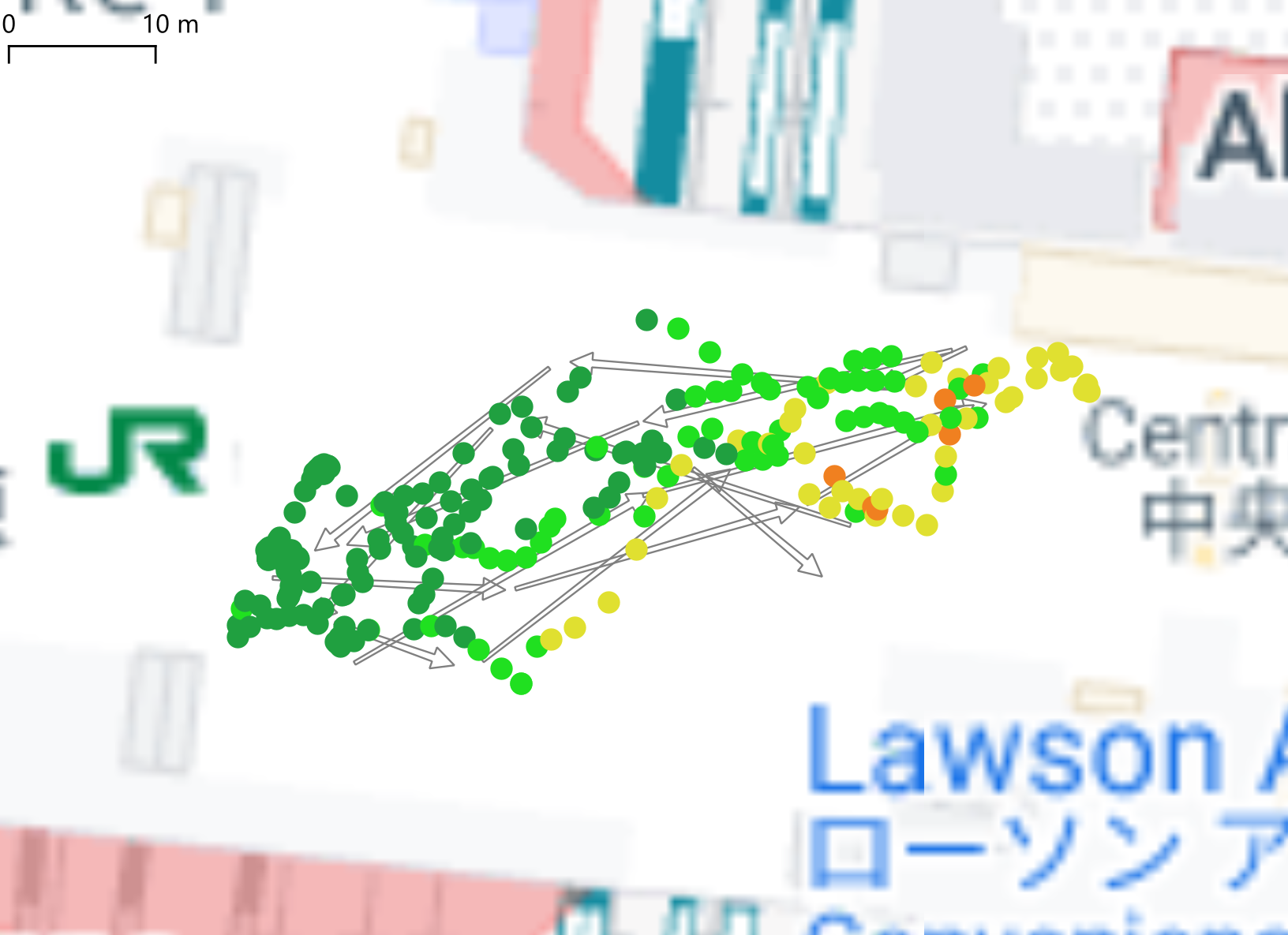}
  \vspace{-1mm}
  \caption{Akihabara Station (JP)}
  \label{fig:AkihabaraWalk}
\end{subfigure}
\begin{subfigure}{.24\textwidth}
  \centering
  \includegraphics[width=0.95\linewidth]{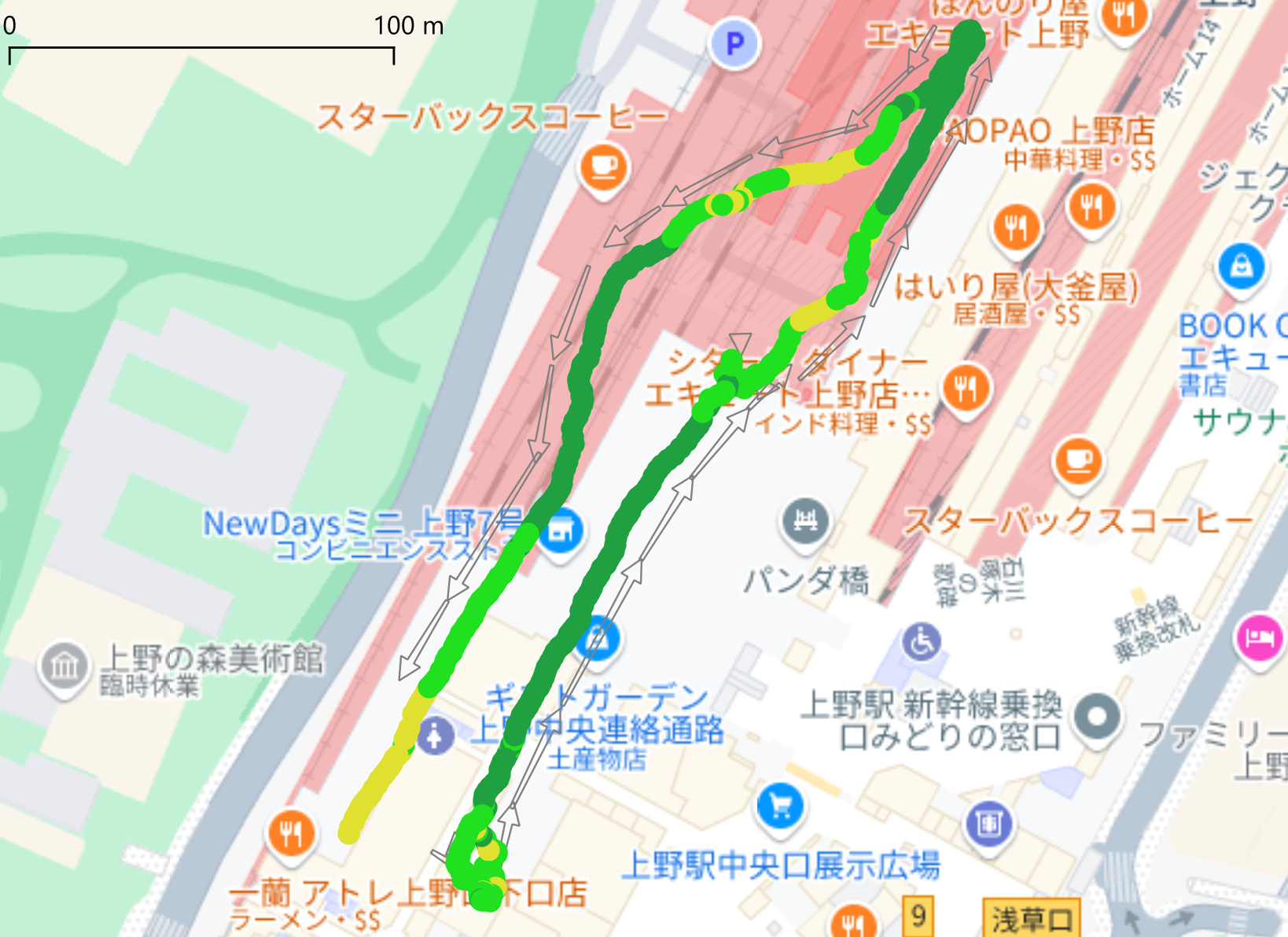}
  \captionsetup{justification=centering}
  \vspace{-1mm}
  \caption{Ueno Station (JP)}
  \label{fig:Ueno}
\end{subfigure}%
\begin{subfigure}{.24\textwidth}
  \centering
  \includegraphics[width=0.95\linewidth]{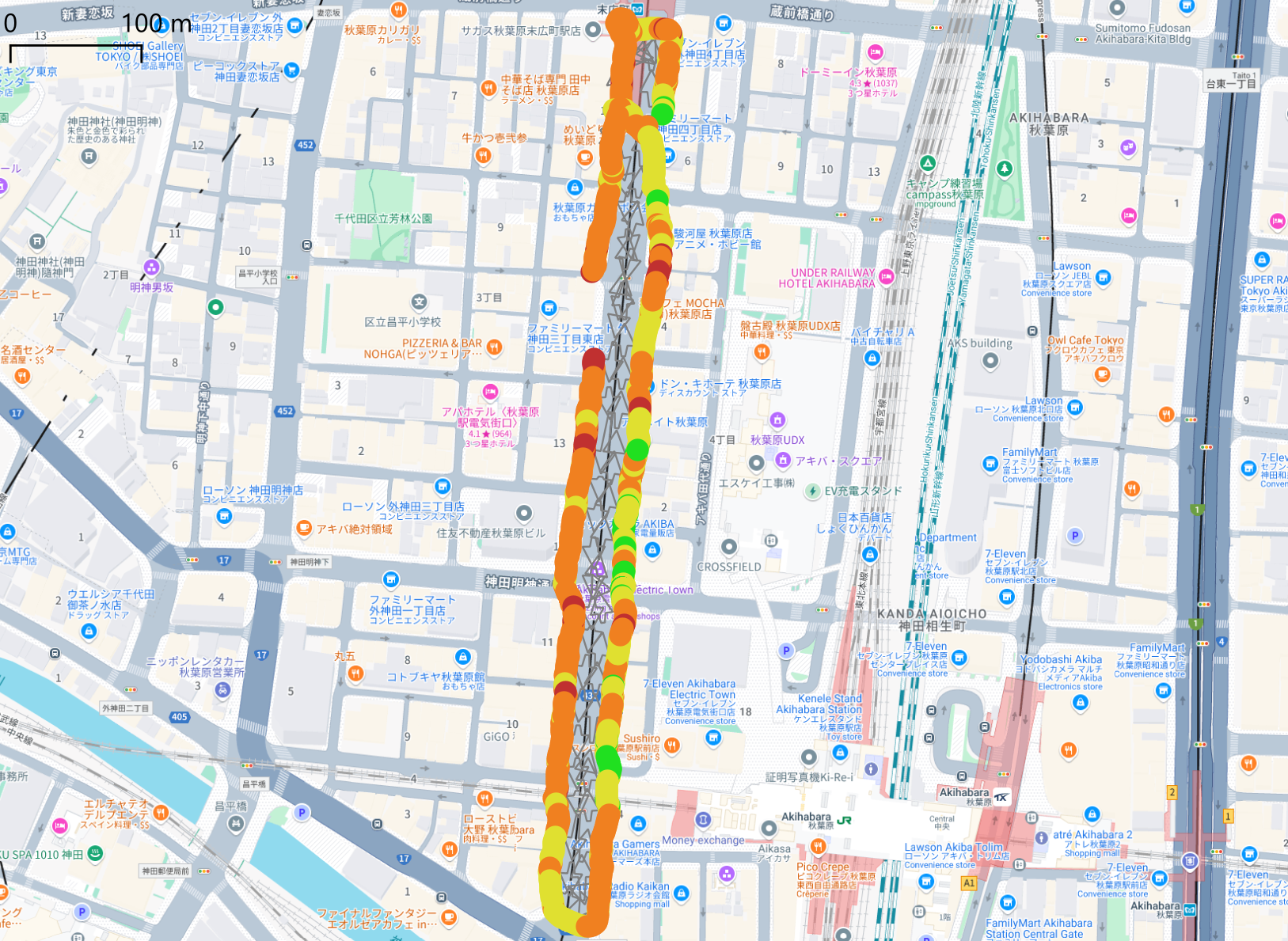}
  \captionsetup{justification=centering}
  \vspace{-1mm}
  \caption{Akihabara Neighborhood (JP)}
  \label{fig:AkihabaraLoop}
\end{subfigure}\\
\begin{subfigure}{.49\textwidth}
  \centering
  \includegraphics[width=0.95\linewidth]{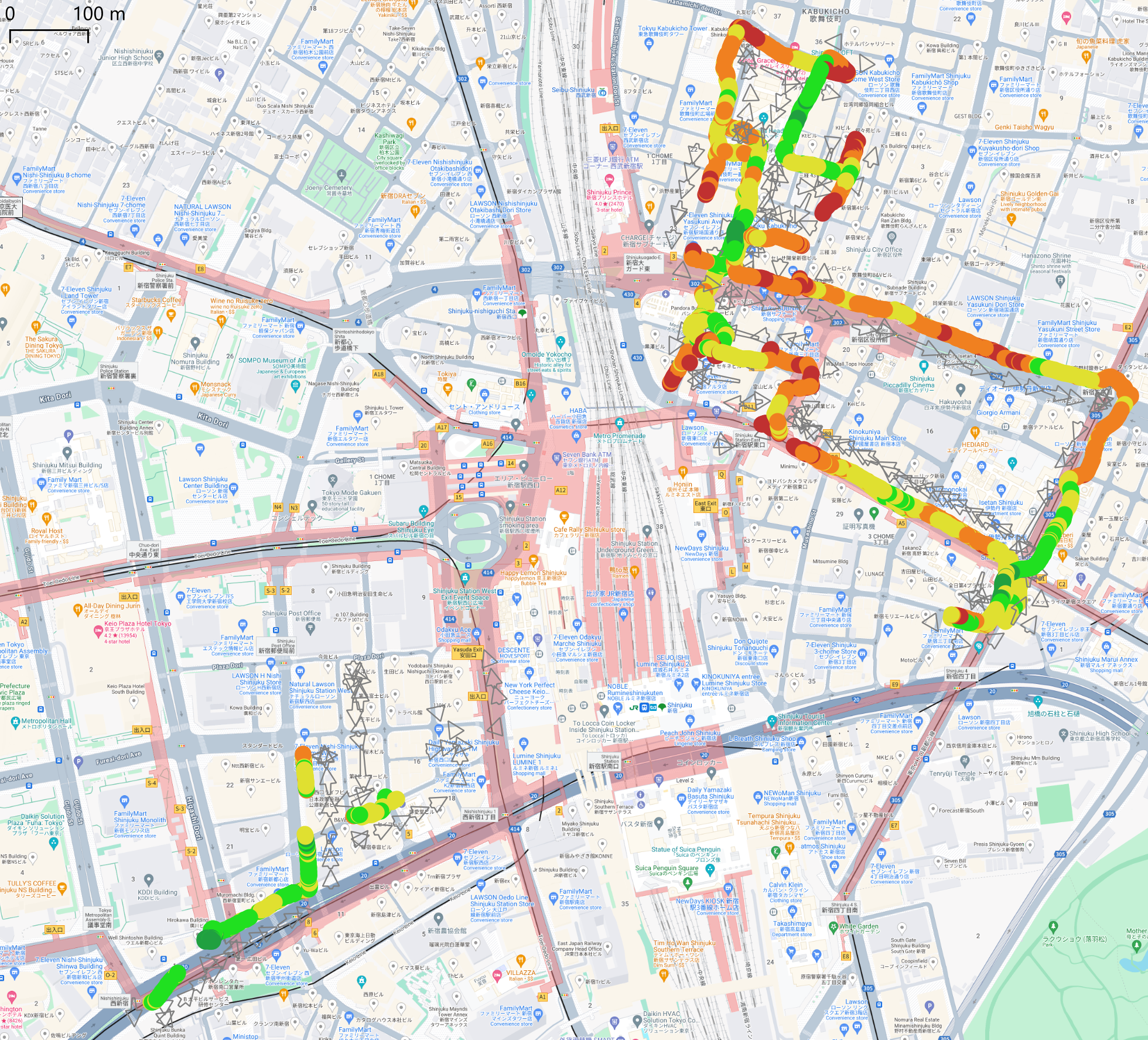}
  \captionsetup{justification=centering}
  \vspace{-1mm}
  \caption{West Shinjuku (Left) and Kabukicho (Right) (JP)}
  \label{fig:Kabuki}
\end{subfigure}
\vspace{-5mm}
\caption{Test Route of each experiment. Color legends represent 5G NR SS-RSRP.}
\label{fig:MobilityMap}
\vspace{-7mm}
\end{figure}

For head-to-head experiments, five locations choose for the experiments are BTS Siam Sky Bridge (Thailand), Sidewalk in front of Waseda University (Japan), Tokyo Station Square (Japan), Akihabara Station (Japan), and Ueno Station (Japan). On the other hand, the \textit{single trial} locations were the neighborhoods of Akihabara (Japan), Kabukicho (Japan), and West Shinjuku (Japan). It should be noted that only some base stations in Kabukicho has DL-256QAM enabled on FR2, so data will be analyzed separately. All SoftBank base stations in the Akihabara area do not support DL-256QAM on FR2. Moreover, due to UE capability concerns, MNOs around the world usually deploy 5G with a channel bandwidth of 100 MHz per carrier, then utilize Carrier Aggregation to combine multiple carriers to fully utilize the frequency spectrum licensed to the MNO. For this reason, it is common for multiple carriers between two and eight being deployed for the FR2. Since reporting the aggregated throughput would be misleading due to the difference in the aggregated bandwidth, the average was taken across all 100 MHz carriers available and reported as throughput per 100 MHz carrier in the following section. The RF conditions at each location by carrier can be seen in Table \ref{tab_RFCondition} and the mobility map can be seen in Fig. \ref{fig:MobilityMap}. Finally, the head-to-head test cases were outlined in Table \ref{tab_headhead}. \looseness=-1

\begin{figure}[t!]
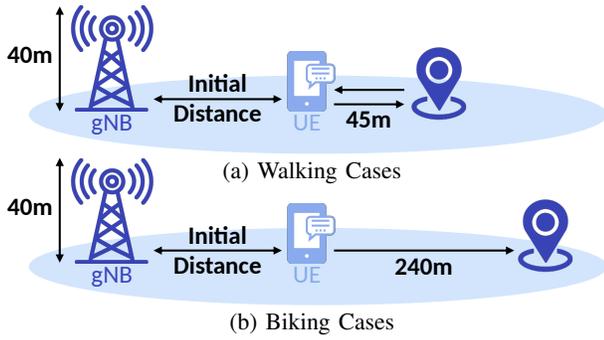


\begin{subfigure}{\linewidth}
\centering\includesvg[width=0.9\linewidth,inkscapelatex=false]{Walking.svg}
  \vspace{-1.3mm}
  \caption{Walking Cases}
  \label{fig-Walking}
  \vspace{-2mm}
\end{subfigure}\\

\begin{subfigure}{\linewidth}
\vspace{-6mm}
\centering\includesvg[width=0.9\linewidth,inkscapelatex=false]{Biking.svg}
  \vspace{-1.3mm}
  \caption{Biking Cases}
  \label{fig-Biking}
\end{subfigure}
\vspace{-5mm}
\caption{Simulation Setup}
\vspace{-3mm}
\label{fig:SimulationSetup}
\end{figure}

\begin{table}[!tbp]
\setstretch{0.8}
\caption{Summary of RF Conditions}
\vspace{-1.5mm}
\centering
\label{tab_RFCondition}
\resizebox{8.7cm}{!}{\begin{tabular}{@{}lccccccc@{}}
\toprule
Location   &Carrier&LTE&LTE&LTE&5G NR&5G NR&5G NR \\
&&RSRP&RSRQ&SINR&RSRP&RSRQ&SINR\\
&&(dBm)&(dB)&(dB)&(dBm)&(dB)&(dB)\\\midrule
Sky Bridge (TH) & AIS & -59.77&-12.71&15.64&-93.92&-10.95&19.26\\
Waseda Sidewalk (JP)&SoftBank&-72.66&-10.96&9.55&-103.18&-11.44&13.20
\\
Tokyo Station Square (JP)&KDDI&-92.33&-12.24&5.38&-72.50&-10.25&31.11\\
&SoftBank&-93.88&-10.69&10.84&-86.50&-11.18&26.87\\
Akihabara Station (JP)&KDDI&-66.92&-10.86&12.09&-77.30&-10.49&30.16\\
Ueno Station (JP)&KDDI&-76.44&-10.33&16.36&-80.18&-10.53&25.64\\\midrule
Kabukicho (qam64) (JP)&SoftBank&-70.57&-11.49&8.32&-99.46&-11.34&15.26\\
Akihabara (qam64) (JP)&SoftBank&-70.58&-11.27&8.82&-100.20&-11.20&15.14\\
Kabukicho (qam256) (JP)&SoftBank&-67.41&-11.57&7.25&-96.34&-11.42&14.17\\
West Shinjuku (qam256) (JP)&KDDI&-71.75&-10.61&14.74&-85.39&-10.57&25.78\\

\bottomrule
\end{tabular}}
\vspace{-4mm}
\end{table}


\begin{table}[!tbp]
\setstretch{0.8}
\vspace{1mm}
\caption{Summary of Head-to-Head Test Cases}
\vspace{-1.5mm}
\centering
\label{tab_headhead}
\resizebox{8.7cm}{!}{\begin{tabular}{@{}lcccl@{}}
\toprule
Case&Carrier&Type&Category&Description\\
\midrule
Sky Bridge (CW)&AIS&NSA&Outdoor&Walk clockwise in loop\\
&&&&(NLOS/LOS, multiple base stations)\\
Sky Bridge (CCW)&AIS&NSA&Outdoor&Walk counterclockwise in loop\\
&&&&(NLOS/LOS, multiple base stations)\\
Sky Bridge (Crowded)&AIS&NSA&Outdoor&Walk clockwise in loop during rush hour \\
&&&&(NLOS/LOS, multiple base stations)\\
Waseda Sidewalk (Walk)&SoftBank&NSA&Outdoor&Walk in loop (NLOS/LOS, single base station)\\
Waseda Sidewalk (Rest)&SoftBank&NSA&Outdoor&Stand in front of base station (single base station)\\
Tokyo Sta. (EN-DC)&KDDI&NSA&Outdoor&Walk in loop (LOS, single base station)\\
Tokyo Sta. (NR-DC)&SoftBank&SA&Outdoor&Walk in loop (LOS, single base station)\\
Akihabara Sta. (Walk)&KDDI&NSA&Indoor&Walk in loop (NLOS/LOS, single base station)\\
Akihabara Sta. (Rest)&KDDI&NSA&Indoor&Stand in front of base station (single base station)\\
Ueno Station&KDDI&NSA&Indoor&Walk in loop (NLOS/LOS, multiple base stations)\\

\bottomrule
\end{tabular}}
\vspace{-5mm}
\end{table}

\subsection{Simulation Setup}

To validate the field test results and explore the effect of 1024QAM on FR2. Simulations were conducted in MATLAB across three Line of Sight (LOS) scenarios: stationary, walking, and biking. The gNB was configured to match AIS' configurations and Huawei HAAU5323 specifications as shown in Table \ref{tab_SimParams}. The walking simulation involves walking 45 meters farther from the base station starting from the initial distance, then returning to the starting point at an average speed of 1.375 m/s (see Fig. \ref{fig-Walking}). On the other hand, for biking simulation, UE is being moved away from the base station to cover the distance of 240 meters at an average speed of 6.7 m/s (see Fig. \ref{fig-Biking}). The simulation time for walking and biking cases is 60 seconds and 30 seconds, respectively. Due to MATLAB lacking the implementation of MCS and CQI Table required for 1024QAM simulation, the required MCS Table 5.1.3.1-4 and CQI Table 5.2.2.1-5 were implemented based on 3GPP TS 38.214 Release 17 \cite{3GPP_38-214} in the file \textit{MACConstants.m} and \textit{nrCQITables.m}, respectively. Then, 1024QAM was added to \textit{getMCSTable.m} and \textit{getCQITable.m} to make MATLAB recognize the newly implemented MCS and CQI tables. Furthermore, \textit{nrGNB.m} was modified to select the appropriate MCS Table for each simulation case. Finally, for the propagation path loss model, the "Urban Micro" or UMi model was implemented based on 3GPP TR 38.901 \cite{3GPP_38-901, 7504435}.

\section{Results and Analysis}
\begin{figure}[t!]
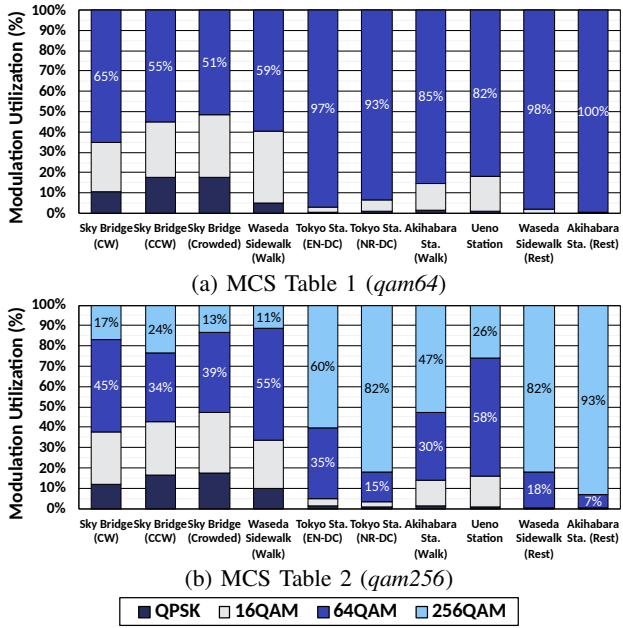


\begin{subfigure}{\linewidth}
\centering
\includesvg[width=0.92\linewidth,inkscapelatex=false]{64QAMUsage.svg}
  \vspace{-1mm}
  \caption{MCS Table 1 (\textit{qam64})}
  \label{fig-64QAMUsage}
\end{subfigure}\\
\begin{subfigure}{\linewidth}
\centering
\includesvg[width=0.92\linewidth,inkscapelatex=false]{256QAMUsage.svg}
  \vspace{-1mm}
  \caption{MCS Table 2 (\textit{qam256})}
  \label{fig-256QAMUsage}
\end{subfigure}
\\
\begin{subfigure}{\linewidth}
\centering
\includesvg[width=0.60\linewidth,inkscapelatex=false]{UtilizationLegend.svg}
\end{subfigure}
\vspace{-5mm}
\caption{Modulation Utilization (\%) for Head-to-Head cases}
\vspace{-5mm}
\label{fig:QAMUse}
\end{figure}
\begin{figure}[t!]
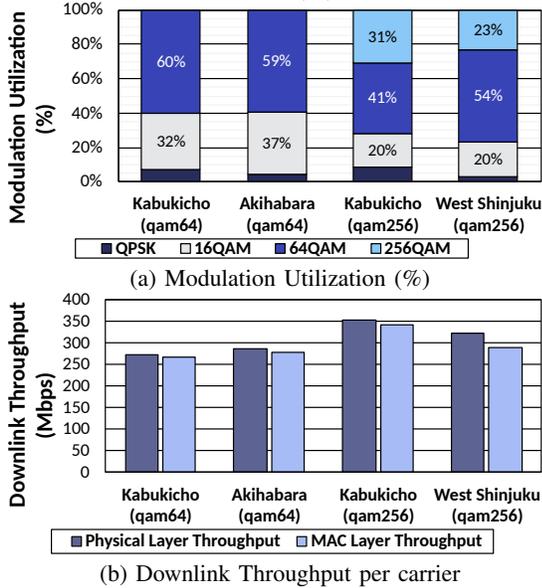


\begin{subfigure}{\linewidth}
\centering
\includesvg[width=0.8\linewidth,inkscapelatex=false]{ModulationNeighborhood.svg}
  \vspace{-1mm}
  \caption{Modulation Utilization (\%)}
  \label{fig-ModulationNeighborhood}
\end{subfigure}\\
\begin{subfigure}{\linewidth}
\centering
\includesvg[width=0.8\linewidth,inkscapelatex=false]{ThptNeighborhood.svg}
  \vspace{-1mm}
  \caption{Downlink Throughput per carrier}
  \label{fig-ThptNeighborhood}
  \vspace{0.5mm}
\end{subfigure}

\vspace{-1.5mm}
\caption{Walking in Neighborhood Field Test Results}
\label{fig:NeighborhoodResults}
\vspace{-6mm}

\end{figure}
\begin{figure*}[t!]
\vspace{1mm}
\begin{subfigure}[t]{.50\textwidth}
  \includesvg[width=0.98\linewidth,inkscapelatex=false]{Thpt.svg}
  \vspace{-1mm}
  \caption{Average Downlink MAC Throughput (Mbps) across all carriers.}
  \label{fig:Thpt}
\end{subfigure}\hfill
\begin{subfigure}[t]{.50\textwidth}
  \includesvg[width=0.98\linewidth,inkscapelatex=false]{ReTx.svg}
  \vspace{-1mm}
  \caption{Retransmission Rate (\%) by Test Case.}
  \label{fig:ReTx}
\end{subfigure}\\

\begin{subfigure}[t]{\linewidth}
\vspace{-4mm}
\centering\includesvg[width=0.33\linewidth,inkscapelatex=false]{BarChartLegend.svg}
\vspace{1mm}
\end{subfigure}\\
\vspace{-4mm}
\begin{subfigure}[t]{.33\textwidth}
  \includesvg[width=0.98\linewidth,inkscapelatex=false]{SoftBankCurve.svg}
  \vspace{-1mm}
  \caption{5G SS-RSRP vs DL-Thpt: SoftBank}
  \label{fig:SBCurve}
\end{subfigure}\hfill
\begin{subfigure}[t]{.33\textwidth}
  \includesvg[width=0.98\linewidth,inkscapelatex=false]{AISCurve.svg}
  \vspace{-1mm}
  \caption{5G SS-RSRP vs DL-Thpt: AIS}
  \label{fig:AISCurve}
\end{subfigure}\hfill
\begin{subfigure}[t]{.33\textwidth}
  \includesvg[width=0.98\linewidth,inkscapelatex=false]{KDDICurve.svg}
  \vspace{-1mm}
  \caption{5G SS-RSRP vs DL-Thpt: KDDI}
  \label{fig:KDDICurve}
\end{subfigure}\hfill
\\
\begin{subfigure}[t]{\linewidth}
\vspace{4.5mm}
\centering\includesvg[width=0.33\linewidth,inkscapelatex=false]{LineChartLegend.svg}
\vspace{2mm}
\end{subfigure}\\

\setlength{\belowcaptionskip}{-18pt}
\vspace{-3mm}
\caption{Head-to-Head Field Test Results}

\end{figure*}

\begin{figure*}[t!]
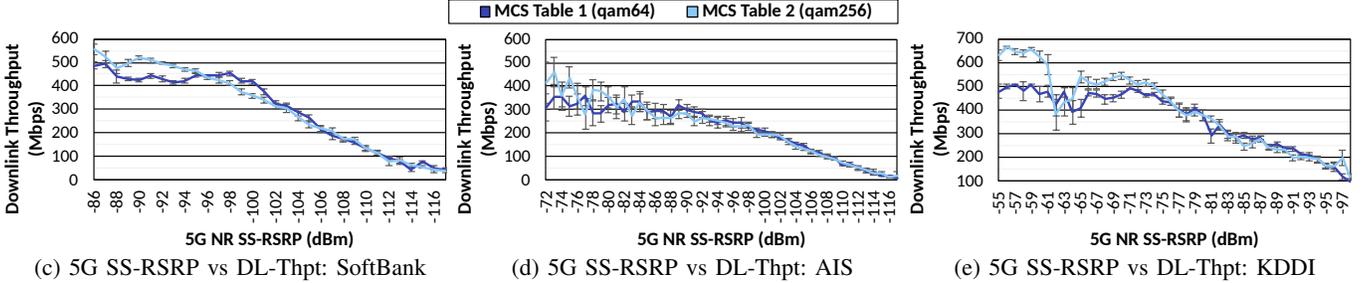
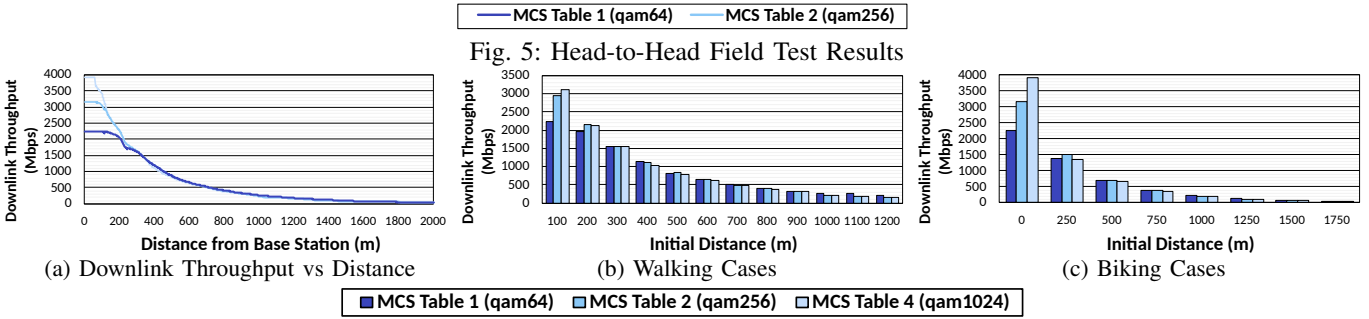

\vspace{1.5mm}
\begin{subfigure}[t]{.33\textwidth}
  \includesvg[width=0.98\linewidth,inkscapelatex=false]{SimulationCurve.svg}
  \vspace{-1.5mm}
  \caption{Downlink Throughput vs Distance}
  \label{fig:SimulationCurve}
\end{subfigure}\hfill
\begin{subfigure}[t]{.33\textwidth}
  \includesvg[width=0.98\linewidth,inkscapelatex=false]{WalkChart.svg}
  \vspace{-1.5mm}
  \caption{Walking Cases}
  \label{fig:WalkChart}
\end{subfigure}\hfill
\begin{subfigure}[t]{.33\textwidth}
  \includesvg[width=0.98\linewidth,inkscapelatex=false]{BikeChart.svg}
  \vspace{-1.5mm}
  \caption{Biking Cases}
  \label{fig:BikeChart}
\end{subfigure}\hfill

\begin{subfigure}[t]{\linewidth}
\vspace{0.5mm}
\centering\includesvg[width=0.5\linewidth,inkscapelatex=false]{SimulationLegend.svg}
\vspace{1mm}
\end{subfigure}\\

\setlength{\belowcaptionskip}{-18pt}
\vspace{-2.5mm}
\caption{Simulation Results}
\label{fig:SimulationResults}
\vspace{-0.8mm}
\end{figure*}

\subsection{Field Test Results}

Fig. \ref{fig:QAMUse} shows the modulation utilization percentage for each test case both when using Modulation Coding Scheme Table 1 (\textit{qam64}) and Table 2 (\textit{qam256}). The results show that when MCS Table 2 was used, an average of 35\% of resource blocks were being modulated using 256QAM instead of 64QAM across all mobility cases, while the amount of 16QAM and QPSK utilization remains similar. Which demonstrates that despite the huge path loss in the FR2, the real-world channel quality is somewhat sufficient to allow partial utilization of 256QAM modulation. However, without UE mobility, a significantly higher utilization of 256QAM can be achieved, with an average of 88\% of resource blocks being modulated using 256QAM, which would imply a significant improvement in spectral efficiency. Unfortunately, Fig. \ref{fig:ReTx}, which compares the retransmission rate between the MCS Table 1 and 2, shows that the retransmission rate increased significantly across all cases by varying degrees, which outweighs the benefits of MCS Table 2 of introducing the 256QAM modulation in the first place. This is reflected in the Downlink Throughput per 100 MHz carrier results (see Fig. \ref{fig:Thpt}), which shows that MCS Table 2 yields an average of 5.8\% improvement across all test cases, with some cases performing better when using the conventional MCS Table 1. While using MCS Table 2 indeed improved the throughput slightly, it comes at the cost of high retransmission rates due to the increase in PDSCH NACK responses, which will negatively impact the latency because the next PDSCH packet will not be decoded until there is an ACK response. Therefore, confirming that for latency-sensitive applications, using higher modulation should be avoided as much as possible, as long as the throughput is sufficient.

Since it was found that MCS Table 2 yielded an improvement under good channel conditions, while may cause throughput deficit under average signal conditions, it is necessary to understand the scenarios that MCS Table 2 and 256QAM are useful to only use them in appropriate scenarios. Fig. \ref{fig:SBCurve} shows the average throughput on SoftBank at each RSRP level, with 95\% confidence intervals displayed for MCS Table 1 and 2. The graph indicates that MCS Table 2 provides a modest improvement in performance under sufficiently good channel conditions. Which in this case, is RSRP above -94 dBm. At approximately -95 dBm, the average throughput for MCS Table 1 and 2 converges, marking a crossover point where the performance gain of MCS Table 2 diminishes. Below this threshold, MCS Table 1 offers a performance advantage, up until approximately -102 dBm where below this threshold, the throughput performance of both tables converges, and performance differences become negligible. Moving on, a similar relationship is also observed on AIS (see Fig. \ref{fig:AISCurve}) where a marginal performance improvement is observed on MCS Table 2 under strong signal conditions. However, the throughput was unstable, and eventually converged, becoming similar to the MCS Table 1 just like before. Lastly, the results on KDDI (see Fig. \ref{fig:KDDICurve}) showed that under very strong signal conditions, the utilization of 256QAM can improve the throughput by over 30\%, which is very close to the theoretical value. However, this was observed at RSRP above -60dBm, which cannot be achieved in most use cases, and the percentage gain begins to diminish rapidly below this threshold. At around -70 dBm, MCS Table 2 now only increases the throughput by approximately 20\%, respective to the MCS Table 1 configuration. This is because although 256QAM is still being utilized at this level for the MCS Table 2 configuration, it uses a lower code rate, whereas the MCS Table 1 configuration uses 64QAM Modulation but at a higher code rate. \looseness=-1

For the field test results from walking in the neighborhoods (see Fig. \ref{fig:NeighborhoodResults}), where the UE used for experiments was held as if used by an end user. In these less controlled scenarios, the average utilization of 256QAM modulation dropped to 27\%. While a substantial MAC layer throughput improvement of 28.5\% was observed in Kabukicho due to high mmWave cell density, the West Shinjuku area, which has DL-256QAM enabled, and the Akihabara area, which only supports DL-64QAM, showed only a 4\% improvement in MAC layer throughput, despite an 11\% improvement in physical layer throughput. This suggests that the spectral efficiency gain from 256QAM is offset by retransmissions, resulting in negligible improvement in real-world performance. Therefore, MNOs should carefully consider FR2 cell density in target areas before enabling DL-256QAM on FR2 to save licensing costs and avoid negatively impacting end users' QoE.

Overall, similar conclusions can be drawn from all test scenarios across the three networks. The utilization of 256QAM consistently leads to increased retransmissions, which, in turn, impacts latency in all scenarios while providing only marginal performance gains. In real-world conditions, achieving sufficient signal quality for 256QAM is unlikely due to rapid path loss on FR2, and when 256QAM was used, it only provided marginal improvements in most scenarios, particularly during mobility cases. Therefore, the results suggest that for latency-sensitive applications such as URLLC, using MCS Table 2 with 256QAM modulation on FR2 is not recommended. For typical enhanced mobile broadband (eMBB) applications that can tolerate higher latency, the increased retransmissions associated with 256QAM are less critical. Under ideal conditions, 256QAM can increase user-perceived throughput by up to 30\%; however, under average signal conditions, using MCS Table 2 may result in worse throughput compared to MCS Table 1, effectively negating any capacity improvements. In a real network with multiple UEs experiencing varying signal qualities, this would likely cancel out the overall benefit of 256QAM. \looseness=-1

\subsection{Simulation Results}

Fig. \ref{fig:SimulationResults} shows the MATLAB Downlink Throughput vs Distance simulation results for stationary, walking, and biking mobility scenarios. The throughput results for MCS Table 1 (\textit{qam64}), 2 (\textit{qam256}), and 4 (\textit{qam1024}) demonstrate that throughput decreases sharply with distance from the base station due to high path loss experienced in FR2. Initially, MCS Table 4 achieves the highest throughput at close range, but its performance drops rapidly within the first 200 meters, quickly converging with that of MCS Tables 1 and 2. This convergence behavior is consistent with the findings from the real-world test results. This indicates that MCS tables with higher modulation are highly sensitive to signal quality and only effective at very close proximity to the base station. It was also found that the MCS Table 2 modulation performed better than MCS Table 4 at 250 meters for the biking mobility scenario, suggesting the need for more intermediate MCS steps. Additionally, similar to the real-world results, MCS Table 2 provides a modest initial throughput advantage over MCS Table 1, but this benefit rapidly diminishes and converges with MCS Table 1. On the other hand, MCS Table 1 demonstrates a more gradual decline in throughput and maintains greater stability over longer distances, ultimately achieving comparable or even slightly higher throughput than MCS Tables with 256QAM and 1024QAM modulation at extended distances. These results suggest that while higher-order QAM schemes offer performance benefits in short-range, line-of-sight conditions, their advantages are severely limited by mmWave’s rapid path loss, making MCS Table 1 a more reliable option at greater distances. 
\looseness=-1

It should be noted that these limitations arise due to the limited number of entries allowed in MCS tables standardized in the 3GPP standard. Each MCS table supports a maximum of 31 entries, requiring tables with higher modulation indexes to take the compromise by having to remove some intermediate MCS entries for lower modulation schemes. This affects the link adaptation ability at mid-cell and cell edge, which may result in MCS tables with higher modulation index to under perform in such scenarios. Some RAN manufacturers have implemented dynamic MCS table switching based on signal conditions to address this issue, but a more effective solution would be to increase the number of allowed entries per MCS table, enabling better link adaptation across all signal conditions. \looseness=-1

\section{Conclusions and Future Work}

The purpose of this study is to evaluate the performance of 256QAM on FR2 in KDDI and SoftBank networks in Japan, and the AIS network in Thailand. Across all three networks, it was found that 256QAM utilization on FR2 is significantly limited due to heavy path loss on FR2 combined with the high MCS requirements for 256QAM. MATLAB simulations on FR2 also suggest that 256QAM utilization will be restricted due to rapid throughput decline over distance. A reasonable performance gain was observed in stationary scenarios, highlighting the impact of mobility on modulation performance in FR2. This suggested that MCS tables with higher modulation schemes such as Table 2 (\textit{qam256}) and Table 4 (\textit{qam1024}) could be useful in improving users' throughput in stationary use cases such as Fixed Wireless Access (FWA) deployments. However, with mobility, the performance gain of 256QAM diminishes significantly, providing only marginal improvements over 64QAM and greatly increasing retransmission rates due to a higher probability of NACK responses. This increase in retransmissions negatively impacts latency performance, making 256QAM unsuitable for latency-sensitive applications such as URLLC; therefore, for these applications, the use of MCS Table 2 is not recommended. The increased retransmissions also limit the potential throughput gain of 256QAM and worsen the latency. Additionally, under average signal conditions, using MCS Table 1 may yield better throughput and stability, effectively canceling out the capacity gains of MCS Table 2 in real network environments where multiple UEs experience varying signal conditions. This suggests that MCS Table 2 could be optimized, and future research will explore custom MCS tables through simulations to address the capacity gain limitations observed in this study.

\section*{Acknowledgement}

This paper is supported by the Ministry of Internal Affairs and Communications (MIC) Project for Efficient Frequency Utilization Toward Wireless IP Multicasting. Additionally, the authors would like to express their gratitude to \textbf{PEI Xiaohong} of \textit{Qtrun Technologies} for providing \textit{Network Signal Guru (NSG)} and \textit{AirScreen}, the cellular network drive test software used for result collection and analysis in this research. Finally, this paper is inspired by \textbf{Furina}, the legendary actress of \textit{Fontaine} who makes all the \textit{Teyvat} her stage.






%

\setstretch{0.9}
\renewcommand{\IEEEbibitemsep}{0pt plus 0.5pt}
\makeatletter
\IEEEtriggercmd{\looseness=-1}
\makeatother
\IEEEtriggeratref{1}
\Urlmuskip=0mu plus 1mu\relax

\bibliographystyle{IEEEtran}

\bibliography{IEEEabrv,bstcontrol,b_reference}

\end{document}